\documentclass[nofootinbib,prd,twocolumn,showpacs,showkeys,preprintnumbers]{revtex4-1}
\usepackage{hyperref,amssymb,amsmath,mathrsfs,bm,graphicx}
\begin{document}
\title {The hyperbolically symmetric black hole}
\author{L. Herrera}
\email{lherrera@usal.es}
\affiliation{Instituto Universitario de F\'isica
Fundamental y Matem\'aticas, Universidad de Salamanca, Salamanca 37007, Spain.}
\author{L. Witten}
\email{lwittenw@gmail.com.}
\affiliation{ Independent Researcher, Ringhouse-180 E 1. Jefferson Road, Rockville, MD 20847,  USA}

\date{\today}
\begin{abstract}
We describe some properties  of the hyperbolically symmetric black hole (hereafter referred to as the $HSBH$) proposed a few years ago. We start by explaining the main motivation behind such an idea, and 
	we determine the main differences between  this scenario and   the classical black hole (hereafter referred to as the $CBH$) scenario. Particularly important are  the facts  that, in the $HSBH$ scenario,  (i) test particles in the region inside the horizon experience a repulsive force that prevents them from reaching the center,  (ii)~test particles may cross the horizon outward  only along the symmetry axis, and (iii) the spacetime within the horizon is static but not spherically symmetric. Next, we examine the differences between the two models of black holes in light of the Landauer principle and the Hawking results on the eventual evaporation of the black hole and the paradox resulting thereof. Finally, we explore what observational signature  could be invoked to confirm or dismiss the model.

\end{abstract}
\date{\today}
\maketitle
\begin{quote}
``{\sc L'interviewer.} - mais croyez-vous que vous devez \^{e}tre d'accord avec ce que la plupart de gens autour de vous pensent?''
\\
``{\sc Ruth.} - eh bien, c'est-\`{a}-dire que, quand je ne le suis pas, je me retrouve toujours \`{a} l'h\^{o}pital...''\\
{\sl R. D. Laing, A. Esterson,} {\it L'\'{e}quilibre mental, la folie et la famille,} {\sl Ed. Maspero (Paris, 1971).\\} 
\end{quote}

\section{Why the {HSBH} Model?}
Before  contrasting the $HSBH$ with the $CBH$, we first present the general arguments leading us to propose the $HSBH$ model \cite{1hw}.

As is well known in the $CBH$ scenario, the spacetime outside the horizon is described by the Schwarzschild solution \cite{SC}, whose line element in polar coordinates reads for 
$R >2M$ (in relativistic units and with signature $+2$) as follows 
:
\begin{eqnarray}
ds^2&=&-\left(1-\frac{2M}{R}\right)dt^2+\frac{dR^2}{\left(1-\frac{2M}{R}\right)}+R^2d\Omega^2, \nonumber \\ d\Omega^2&=&d\theta^2+\sin^2 \theta d\phi^2,
\label{w2b}
\end{eqnarray}
where $M$, which measures the total mass--energy of the source, is the only parameter of the solution. 

Such a metric being  static and spherically symmetric  admits the four Killing vectors
\begin{eqnarray}
{\bf{X}_{\bf{(0})}} = \partial _{\bf{t}}, \quad {\bf{X}_{(2)}}=-\cos \phi \partial_{\theta}+\cot\theta \sin\phi \partial_{\phi},\nonumber \\
{\bf X_{(1)}}=\partial_{\phi}, \quad {\bf X_{(3)}}=\sin \phi \partial_{\theta}+\cot\theta \cos\phi \partial_{\phi}.
\label{2cmh}
\end{eqnarray}
 
Also, as is well known,  the singularity appearing in (\ref{w2b}) for $R=2M$ is  not a real physical singularity, and the solution may be analytically extended to the whole spacetime  (including the region inside the horizon), as shown in \cite{1c,fin,krus,is}.

However, the fact is that the spacetime within the horizon is necessarily non-static. Maintaining  the static form of the Schwarzschild metric (in the whole spacetime) is incompatible with    the possibility of removing  the coordinate singularity appearing on the horizon in the line element  \cite{rosen}. This is related to the fact that static observers cannot be defined inside the horizon (see \cite{Rin,Caroll} for a discussion on this point). A simple way to see why this comes about  consists of finding out  the form of the null cone at any point inside the horizon by solving the equation
\begin{equation}
ds=0.
\label{ng}
\end{equation}

The solution to the above equation shows that all the null rays generating the null cone (inside the horizon) converge to the center of symmetry, implying that anything within the null cone (including the massless particles along the null cone border) should reach the center (see \cite{Rin,Caroll,pap} for a discussion on this point).

Although most of our colleagues are satisfied with this picture, we are not. Our concern is generated by the intuitive idea that any dynamical process should relax to an equilibrium state after some finite proper time. In such a case, we should be able to provide a global static description of the spacetime.

To the best of our knowledge, the first to call attention to this issue was Rosen \cite{rosen}, who proposed excluding the region $R<2M$ because, in Rosen's words, ``\dots the surface $R=2M$ represents the boundary of physical space and should be regarded as an impenetrable barrier for particles and light rays'' (see page 233 in \cite{rosen}).

Sharing Rosen's concern about the physical properties of  the region $R<2M$, we proposed in  \cite{1hw} an alternative way to overcome the above-mentioned situation.

Thus, we proposed describing the region $R >2M$ via the Schwarzschild solution (\ref{w2b}); however, instead of excluding the region $R <2 M$, we assume that the spacetime in that region is defined  by the line element  (with signature $-2$) 
\begin{eqnarray}
ds^2&=&\left(\frac{2M}{R}-1\right)dt^2-\frac{dR^2}{\left(\frac{2M}{R}-1\right)}-R^2d\Omega^2, \nonumber \\ d\Omega^2&=&d\theta^2+\sinh^2 \theta d\phi^2.
\label{w3c}
\end{eqnarray}

The above metric is static; therefore, it  admits the time-like Killing vector  $\bf{X }_{(\mathbf{0})}$,  and,  furthermore, it admits  the three Killing vectors
\begin{eqnarray}
{\bf Y_{(2)}}=-\cos \phi \partial_{\theta}+\coth\theta \sin\phi \partial_{\phi},\nonumber \\
{\bf Y_{(1)}}=\partial_{\phi}, \quad {\bf Y_{(3)}}=\sin \phi \partial_{\theta}+\coth\theta \cos\phi \partial_{\phi},
\label{2cmhy}
\end{eqnarray}
implying that it is not spherically symmetric but hyperbolically symmetric.

The above three Killing vectors (\ref{2cmhy})  define the hyperbolical symmetry.
For applications of this kind of symmetry,  see \cite{Ha,ellis,1n,Ga,Ri,Ka,Ma,mimc2,2nc,pak1,hd,st1,hd2,pak3,Lim,hn1,hn6,hn7,hn11,hn11b,hn11c,hn12,hn13} and the references therein.

The situation may be summarized as follows: We have already seen that, when extending the Schwarzschild metric to the region inner to the horizon, which means keeping the spherical symmetry, one should abandon staticity; this is the picture for the $CBH$. However, we wish to keep staticity  within the horizon. Accordingly, we propose abandoning sphericity, which leads to the $HSBH$ picture. 

A short comment is in order at this point: In \cite{1hw}, we arrive at (\ref{w3c}) from (\ref{w2b}) through the transformation $\theta\rightarrow i\theta$. In doing so, the signature of the resulting line element (\ref{w3c}) is different from the signature of (\ref{w2b}). However, obviously, (\ref{w3c}) is a static solution to  the vacuum Einstein equations independently on its signature. In other words, the transformation mentioned above is just a technicality, and we may assume the  region inside the horizon to be described by a hyperbolically symmetric metric with the same signature as the Schwarzschild line element describing the region outside the horizon. Thus, we do have a change in symmetry across the horizon but not necessarily a change in its signature. 
 \subsection{Geodesics in HSBH}

In \cite{2nc}, a general study of  geodesics in the spacetime described by (\ref{w3c}) is presented (see also \cite{Lim}), leading to some interesting conclusions that highlight great differences between the behavior of  a test  particle in the $HSBH$ and the results emerging from the $CBH$,  particularly the following:
\begin{itemize}
\item  Inside the region $R<2M$, the gravitational force appears to be repulsive.
\item As a consequence of the above, test particles never reach the center.
\item Unlike the $CBH$, test particles can cross the horizon outward, but only along the $\theta=0$ axis.
\end{itemize}

Based on the last point above, it could be (rightly) argued that the object described by the $HSBH$ is not really ``black'' since matter may cross the horizon outwardly. However, in order  to keep the link between  the two scenarios (the $HSBH$ and $CBH$), we decide to keep the term ``black''.

The results above are supported by the expression of   the four-acceleration ($a^\mu$) of  a static observer in the frame of (\ref{w3c}). 
Indeed,  $a^\mu$  represents  the inertial radial acceleration that should be applied  to the frame in order to keep the frame static by opposing the gravitational acceleration exerted on it.

Next,  for  a static observer, the  four-velocity $U^\mu$ is proportional to the Killing time-like vector \cite{Caroll}. Then, for (\ref{w3c}), we obtain 
\begin{equation}
U^\mu=\left[\frac{1}{(\frac{2M}{r}-1)^{1/2}}, 0, 0, 0 \right],
\label{1a}
\end{equation}
producing  for the four-acceleration $a^\mu\equiv U^\beta U^\mu_{;\beta}$ within the  region inner to the horizon

\begin{equation}\label{3a}
a^\mu=\left[0,-\frac{M }{r^2}, 0, 0\right],
\end{equation}
whereas for the region outside the horizon, defined  by  (\ref{w2b}), the four-velocity vector of a static observer reads 
\begin{equation}
U^\mu=\left[\frac{1}{(1-\frac{2M}{r})^{1/2}}, 0, 0, 0 \right],
\label{1a}
\end{equation}
producing for the four-acceleration
\begin{equation}\label{3ab}
a^\mu=\left[0,\frac{M}{r^2}, 0, 0\right].
\end{equation}

From a simple inspection of  (\ref{3ab}), we see that  the gravitational force  in the  region outer to the horizon is attractive, as it follows from  the  positive  value of the  acceleration (directed radially) in this region, while it is repulsive   in the region inner to  the horizon, as it follows from (\ref{3a}).   The repulsive nature of gravitation  in the spacetime described by (\ref{w3c}) is characteristic of hyperbolical spacetimes and is at the origin of the peculiarities of the orbits within the horizon  in the $HSBH$.

\subsection{Flow of Information and Landauer Principle}
The Landauer principle \cite{Lan} asserts that, in the process of erasing one bit of information, some energy must be dissipated, a lower bound of which   is given by
\begin{equation}
\bigtriangleup E=kT \ln2,
\label{lan1}
\end{equation}
where $k$ is the Boltzmann constant, and $T$ denotes the  temperature of the environment (see also \cite{3,7}). 

Its relevance stems from the fact   that  it leads in a natural way to an ``informational'' reformulation of thermodynamics, which, in turn, allows for establishing a link between information theory and different branches of  science \cite{B, Plenio,   bais, BII}.

The link between  the Landauer principle and general relativity is discussed in detail in \cite{nnref2}. Here, we resort to some results found in that reference to carry on our discussion on the differences between the $HSBH$ and $CBH$.

Thus,   a mass given by 
\begin{equation}
M_{bit}=\frac{kT}{c^2}\ln2,
\label{1nN}
\end{equation}
was  assigned to any  bit of information, where $c$ is the speed of light \cite{3mn}. 

This issue (the mass of information)  has been discussed by several authors (see \cite{1bb, 2bb, vop} and the references therein).

Our goal in this subsection  consists of contrasting the behavior of the information flow across the horizon in the gravitational field of a static  mass in the $HSBH$ and $CBH$ scenarios. To this end, first, we need an expression for the Landauer principle in the presence of a  gravitational field.

The presence of a gravitational field  in the weak field approximation produces a change in the total amount of minimal dissipated energy, which is  given by \cite{pla} 
\begin{equation} 
\bigtriangleup E=\frac{kT\left(1+\frac{\phi}{c^2}\right)}{c^2}\ln2,
\label{1nN}
\end{equation}
where $\phi$ denotes the Newtonian (negative) gravitational potential. Thus, in the presence of a gravitational field, the Landauer principle is modified by replacing the temperature $T$ by the Tolman temperature  \cite{Tol}, which, in the weak field limit, reads  $T\left(1+\frac{\phi}{c^2}\right)$. 

Let us recall  that, in presence of a gravitational field, thermodynamic equilibrium is ensured by the condition that the Tolman temperature is constant. This result, which is a consequence of  the inertia of thermal energy, emerges naturally  in the relativistic transport equations proposed in \cite{E, L, 21T}, and it should hold in any  physically admissible,  relativistic transport equation.

The generalization of (\ref{lan1})  for  a gravitational field of an arbitrary intensity  reads 
\begin{equation}
\Delta E=kT \sqrt{\vert g_{tt}\vert}\ln2,
\label{3}
\end{equation}
whereas  the expression for the mass of a bit becomes
\begin{equation}
M_{bit}=\frac{kT\sqrt{\vert g_{tt}\vert}}{c^2}\ln2,
\label{3nl}
\end{equation}
where $g_{tt}$ is the $tt$ component of the metric tensor, and the fact is used that the  Tolman temperature is defined by $T \sqrt{\vert g_{tt}\vert}$, which, in the weak field limit, becomes $T(1+\frac{\phi}{c^2})$. 

Let us now consider the $CBH$ scenario.

In this case,  if we admit  that a bit of information is endowed with a mass according to~(\ref{3nl}), then, since nothing can cross the horizon outwardly, it follows  that no information can cross  the horizon outwardly. 
This  conclusion agrees with the results obtained by Hawking  about the radiation of a $CBH$ due to  quantum effects  \cite{H1} and its eventual  evaporation, as well as with the fact that such radiation is completely thermal (i.e.,  it conveys no information) \cite{H2}),  which leads to a  contradiction known as  the information loss paradox.

Indeed, the  Hawking result  implies  that a pure  quantum state  evolves into a mixed state (the thermal radiation), which, of course, contradicts  the principle of unitary evolution. Since the Hawking radiation is thermal,  it appears that all information about the collapsing object is  lost forever.

In spite of the intense research work devoted to this issue in recent decades, no satisfactory resolution of this quandary has yet been proposed.

Let  us next consider the $HSBH$ scenario.

As mentioned before, in such a case, the crossing of massive particles through the horizon outwardly is allowed  along the $\theta=0$ axis, 
thereby implying the existence of  a flux of information from the inside of the horizon to the outside. Thus, in this scenario, no information loss paradox appears, since even if the Hawking radiation is thermal, information may leave the collapsed object.
 
 Furthermore, in this latter  picture, the energy flow along the axis $\theta=0$ associated with the change in information within the horizon  could be, in principle, observable.

Thus, invoking the Landauer principle in the study of the global picture of the Schwarzschild black hole may lead to observable consequences, which could help to elucidate the quandary about the real nature of the Schwarzschild black hole.

\section{Observational Evidences}
Finally, we would like to mention some observational  projects that could be determinant to dismiss or confirm the $HSBH$ scenario.

The most important of  these endeavors seems to be a study based on the information provided by the Event Horizon Telescope (EHT) Collaboration \cite{et1, et2,ncr}, which provides observations of shadow images of the gravitationally collapsed objects at the center of the elliptical galaxy $M87$  and at the center of the Milky Way Galaxy.  Such observations are expected to provide important data on strong fields \cite{et3, et4, et5}, which could   be used to obtain constraints for the parameters of the solutions describing  the geometry surrounding  compact objects, e.g.,   black hole spacetimes in modified and alternative theories of gravity~\cite{ex1, ex2, ex3, ex4, ex5}, naked singularities, and classical GR black holes with hair or immersed in matter fields \cite{ex9, ex10, ex11, ex12}.

It should be noted that, although the region exterior to the horizon is described by the usual  Schwarzschild metric, we might expect some imprints on the shadows from the material ejected along the symmetry axis in the $HSBH$.

Another possible source of  information to support the $HSBH$ picture could be the observation and modeling of 
 extragalactic relativistic jets.
 
These are  highly energetic phenomena that have been observed in many systems (see~\cite{Bladford, Margon, Sams,blan} and the references therein), usually associated with the presence of a compact object. One of their characteristic properties is a high degree of collimation (besides the extremely high energies involved).  

So far, no consensus has been reached concerning the basic mechanism explaining these two features of jets (collimation and high energies); therefore, it is legitimate  to  speculate that the $HSBH$  could be considered a possible engine behind the jets.

 Indeed, let us recall that, in the $HSBH$ scenario, test particles may cross the horizon outward, but only along the $\theta=0$ axis, thereby explaining the collimation. However, as it follows from  (\ref{3a}), the strength of the repulsive gravitational force acting on the particle as $r\rightarrow 0$ increases as $\frac{1}{r^{2}}$, thereby explaining the high energies of the particles bouncing back from regions close to $r=0$.
 
Obviously,  at this level of generality, this is just speculation.  A solid theory for relativistic jets
would require   a much more detailed setup based on astronomical observations.

\section{Discussion and Conclusions}

Above, we exposed the main features of the $HSBH$ and contrasted them with those of the $CBH$. We started by explaining that our main motivation for proposing the $HSBH$ was the need to describe the region interior to the horizon by a static spacetime, which, in turn, forced us to abandon the assumption of spherical symmetry in that region. The analysis of the geodesic structure within the horizon exhibited profound differences from the corresponding structure in the $CBH$. 

As we showed, such differences are related to the fact that the nature of gravity within the horizon happens to be repulsive, as a consequence of which test particles never reach the center. This feature is perhaps the most relevant difference between the  $HSBH$ and $CBH$.

The origin of the repulsive nature of gravitation within the horizon should be found in the presence of negative mass (energy). This is indeed the case, as shown in the studies of fluid distributions endowed with hyperbolical symmetry (see \cite{mimc2, hd, st1, hn11b} for details). 

It is worth mentioning that the existence of negative mass (energy) in gravitational physics has a long and venerable history, at both the classical \cite{bon,bo1,co1,pap,neg,nm,bor} and quantum  levels \cite{nm1, we1, we2, we3, cqg, pav, hao}.

We next invoked the Landauer principle to illustrate the differences between the $HSBH$ and $CBH$ concerning the possible flow of information and the discussion of the Hawking results in both scenarios. It follows  in a natural way that no information loss paradox appears in the $HSBH$.

Finally, we call attention to a new line of investigations involving  observations of shadow images of the gravitationally collapsed outcome, aiming to identify the nature of different compact objects. This could provide arguments to support (or dismiss) the $HSBH$ picture. Also, the idea has been put forward about  the possible use of the $HSBH$ as the engine of extragalactic jets.

Before concluding, we would like to highlight three important issues that we believe deserve  further investigation:
\begin{itemize}
\item The two manifolds describing the inner and outer parts of the horizon do not match smoothly in the Darmois sense \cite{14nn}. This implies that there is a shell on the horizon, whose physical and mathematical properties are described by the Israel conditions~\cite{15nn}. It would be interesting to delve deeper into this issue and find out how these properties affect the $HSBH$ scenario.
\item It would be interesting to find out whether a ``thermodynamic'' approach {\it a la Bekenstein} may also be applied to the $HSBH$.
\item The $HSBH$ scenario implies a change in symmetry across the horizon;  is there any physical explanation for this (e.g., a phase transition)?
\end{itemize}

\end{document}